\documentclass[aps, showpacs, twocolumn, amsmath, amssymb, prb]{revtex4}
\usepackage{graphicx} 
\newcommand{\rn}{\mathcal{R}_n}

\usepackage{color}
\definecolor{Blue}{rgb}{0.3,0.3,0.9}
\usepackage[T1]{fontenc}
\usepackage{ae,aecompl}
\usepackage{pstricks}
\usepackage{pstricks-add}

\begin{document}


\title{Impurity entanglement entropy in Kondo systems from conformal field theory}


\author{Erik Eriksson}
\author{Henrik Johannesson}
\affiliation{Department of Physics, University of Gothenburg, SE 412 96 Gothenburg, Sweden}


\begin{abstract}
The entanglement entropy in Kondo impurity systems is studied analytically using conformal field theory. From the impurity contribution to the scaling corrections of the entanglement entropy we extract information about the screening cloud profile for general non-Fermi-liquid fixed points. By also considering the finite-temperature corrections to scaling of the von Neumann entropy we point out a direct connection between the large-distance screening cloud profile and thermodynamic observables such as the specific heat. 
\end{abstract}



\maketitle


{\em Introduction.}
Entanglement $-$ nonseparability of states $-$ lies at the very heart of quantum theory. It has been recognized as the crucial resource needed for performing quantum computing and teleportation, but has also entered as an important concept in a wide range of fields spanning from black holes to biological systems.\cite{Horodecki} Entanglement entropy, a measure of the entanglement between two parts of a quantum many-body system, has been used as a theoretical tool to study quantum phase transitions as well as characterizing topological phases of matter\cite{AFOV} and developing numerical algorithms.\cite{VCM} A particularly important result has been for one-dimensional (1D) critical systems, where conformal field theory yields a universal prediction for the scaling of the entanglement entropy.\cite{CC1,CC2} When such a system has a boundary, it provides a framework for the description of universal features of quantum impurity systems.\cite{ALS} Important information can be obtained by studying the subleading corrections to the scaling of the entanglement entropy,\cite{CC3} suggesting a new perspective on the long-standing problem of the evasive ''Kondo screening cloud''.\cite{SCLA}

Regarding the question of measuring entanglement entropy experimentally, it needs to be related to observable quantities. Recent attempts have focused on the connection between entanglement and fluctuations.\cite{KL,SRH} In particular, for critical one-dimensional systems with boundaries there is not only the same logarithmic scaling, but also the corrections to scaling of the entanglement entropy and those of the particle (or spin) fluctuations show interesting similarities.\cite{SRH}

In this paper, we employ boundary conformal field theory (BCFT) to give a unified picture of the entanglement entropy in quantum impurity problems. In particular, we unveil another type of connection between zero-temperature entanglement and fluctuations, this time finite-temperature thermodynamic fluctuations (in the form of the impurity specific heat), at the level of scaling corrections. As we shall argue below, this provides an experimental inroad to the study of Kondo screening clouds. The same framework allows us to extend the Fermi-liquid analysis of S\o rensen {\em et al.}~\cite{SCLA} of the large-distance behavior of the entanglement entropy generated by a Kondo impurity, to the whole range of non-Fermi-liquid fixed points in the large family of Kondo systems.

For the Fermi-liquid case, there is an intuitive picture relating the impurity contribution to the entanglement entropy and the form of the Kondo screening cloud.\cite{SCLA} In a valence-bond basis, the screening cloud can be thought of as the distribution of singlet bonds between the impurity and conduction electrons, hence the valence bond entanglement entropy~\cite{ACLM} provides the connection. Now, there is no such intuitive picture for the non-Fermi-liquid case of the more complex Kondo systems, but the analogy provides a rationale for regarding the impurity entanglement entropy as a good description of the screening cloud profile even when the (partial) screening is carried out by the highly non-trivial objects that make up the non-Fermi-liquid ground state. Importantly, our BCFT result which establishes an exact relation between the impurity contributions to the entanglement and the specific heat in terms of scaling corrections of the von Neumann entropy then shows that there is a direct connection between the large-distance behavior of the screening cloud and an experimental observable.

{\em von Neumann entropy from BCFT.}
Entanglement entropy in a zero-temperature bipartite system is encoded in the von Neumann entropy.  For a critical 1D semi-infinite bipartite system at inverse temperature $\beta$, the scaling of the von Neumann entropy $S(r) = - \mbox{Tr}\hat{\rho}(r) \ln \hat{\rho}(r)$ of the reduced density matrix $\hat{\rho}(r)$ for an interval of length $r$ at the boundary of the system is given by~\cite{CC1,ZBFS}
\begin{equation} \label{S_A}
S(r) = \frac{c}{6} \ln \left[ \frac{\beta}{\epsilon \pi}\sinh \left(\frac{2 \pi r}{\beta} \right) \right] + \ln g + c'  + ... \, .
\end{equation}
Here $c$ is the central charge, $\ln g$ is the boundary entropy,~\cite{AL} $\epsilon$ is a short-distance cutoff, and $c'$ is a non-universal constant. This relates the logarithmic scaling in $r$ of the zero-temperature entanglement entropy to the scaling of the extensive thermodynamic entropy when $r/\beta \to \infty$. We will now see that there is a somewhat similar relation also for the \textit{scaling corrections } of $S(r)$, denoted by "$...$" in Eq.~(\ref{S_A}). These corrections are governed by the irrelevant operators (in the language of the renormalization group) in the BCFT that describes the critical properties of the system, and will have both bulk and boundary contributions. For quantum impurity systems, the fixed-point properties can be described by a BCFT where the impurity has been reduced to a specific boundary condition and boundary operator content.~\cite{Affleck} Therefore, we see from Eq.~(\ref{S_A}) that the impurity contribution $S_{imp}(r)$ to the von Neumann entropy $S(r)$ is
\begin{equation} \label{S_imp}
S_{imp}(r) = \ln g + ... \, .
\end{equation}
Now "$...$" is the boundary contribution to the scaling corrections of $S(r)$, which is governed by the irrelevant \textit{boundary} operators $\{\phi_b\}$ in the theory. 

The von Neumann entropy is calculated as $S(r) = \lim_{n \rightarrow 1} (1-n)^{-1}\ln \mbox{Tr} \hat{\rho}^{n}(r)$. When perturbing the BCFT with an irrelevant boundary operator, $H=H_{{\text BCFT}} + \lambda  \,\phi_b $, the leading correction to 
$S(r)$ is  generically of second order in the scaling field $\lambda$, and can be written as
\begin{equation} \label{int}
\delta S
\!\sim\! \frac{\lambda^2}{4} \! \int^{\beta/2}_{-\beta/2}\! \mathrm{d} \tau_1  \int^{\beta/2}_{-\beta/2} \!\mathrm{d} \tau_2  
 \frac{\left( \left|\frac{ \mathrm{d}z}{\mathrm{d}w}\right|_{w=\tau_1}^{1\!-\!x_b}\! -\! \left|\frac{ \mathrm{d}z}{\mathrm{d}w}\right|_{w=\tau_2}^{1\!-\!x_b}  \right)^{\!2} }{|\frac{\beta}{\pi} \sin( \frac{\pi}{\beta}(\tau_1  - \tau_2))|^{2x_b}}
\end{equation}
Here $|$d$z$/d$w$| comes from the mapping  $z \mapsto w$ that takes the $n$-sheeted Riemann surface $\rn$ in the finite-temperature geometry (representing $\mbox{Tr} \hat{\rho}^{n}$) to the finite-temperature strip $\{ w = \tau + \textrm{i}y \mid -\beta/2 \leq \tau \leq \beta/2, y \geq 0 \}$ in the complex upper half-plane $\mathbb{C}^+$. The cutoff in the integral is given by $ |\tau_1 - \tau_2| \geq \epsilon / | \mathrm{d}z /  \mathrm{d}w|_{w=\tau_1} $. Eq. (\ref{int}), where $x_b$ is the scaling dimension of $\phi_b$, follows from the analysis in Ref.~\onlinecite{EJ}, generalized to finite temperature.  

In the limit $r / \beta \to 0$, we get $|$d$z$/d$w| \propto r \!+\! \mathcal{O}((r / \beta)^3)$. This gives, as expected from the zero-temperature result in Ref.~\onlinecite{EJ}, that $S_{imp} = \ln g + \delta S_{imp}$ where $\delta S_{imp}$ is found from Eq.~(\ref{int}) as
\begin{equation} \label{simp}
\delta S_{imp} \sim \left\{
\begin{array}{ll}
r^{2-2x_b} & \text{if } 1< x_b < 3/2\\
r^{-1} \ln r & \text{if } x_b = 3/2\\
r^{-1} & \text{if } x_b > 3/2 \, ,
\end{array} \right.
\end{equation}
up to terms $\mathcal{O}((r / \beta)^3)$. A marginally irrelevant perturbation generates a leading correction $\sim (\ln \ell )^{-3}$, while an exactly marginal perturbation describes new fixed points and hence changes $S_{imp}$ by a constant. The above corrections are all second-order in $\lambda$; the only operator which gives a non-vanishing first-order correction is the stress-energy tensor which gives a correction\cite{SCLA} $\sim r^{-1}$. 

It is also possible to extract the behavior in the limit $r / \beta \to \infty$ (still at low temperature, i.e.~large $\beta$, but taking $r \gg \beta$). We then get that $|$d$z$/d$w| \propto \beta + \mathcal{O}(\beta^{-1})$, up to terms $\mathcal{O}(e^{-2 \pi r / \beta})$. Changing integration variable to $u = \tan( \pi |\tau_1 - \tau_2 | / \beta)$ and expanding the integrand in Eq.~(\ref{int}) in a power series at the divergence at $u=0$, gives
\begin{eqnarray}
\delta S_{imp} &\sim& \lambda^2 \beta^{5-4x_b} \displaystyle \int^{\beta/2}_{-\beta/2} \mathrm{d} \tau_1  \int^{\infty}_{0} \mathrm{d} u  \Big[ f_1(\tau_1) u^{2-2x_b} \nonumber \\
&&+ (f_2(\tau_1) + \beta f_3(\tau_1)) u^{3-2x_b} + ... \Big] \, , \label{int2}
\end{eqnarray}
where $f_i(\tau_1)$ are regular functions. When $x_b \geq 3/2$, we must use the short-time cutoff, which in the $u$ variable becomes $u \geq\pi \epsilon / (\beta| \mathrm{d}z /  \mathrm{d}w|_{w=\tau_1} ) \sim \epsilon / \beta^2$. Then the leading $\beta$ dependence goes as $\delta S_{imp} \sim \beta^{5-4x_b}\beta^{4x_b-6}=\beta^{-1}$ when $x_b > 3/2$, and $\delta S_{imp} \sim \beta^{5-4x_b} \ln\beta = \beta^{-1}\ln\beta$ when $x_b = 3/2$. When $1< x_b <3/2$, we see from Eq.~(\ref{int2}) that the integral (\ref{int}) converges. Hence the leading $\beta$ dependence comes from the prefactor $\beta^{2-2x_b}$ arising from $|$d$z$/d$w| \propto \beta$. Summarizing, in the limit $r/\beta \to \infty$ we get $S_{imp} = \ln g + \delta S_{imp}$, with
\begin{equation} \label{simpbeta}
\delta S_{imp} \sim \left\{
\begin{array}{ll}
\beta^{2-2x_b} & \text{if } 1< x_b < 3/2\\
\beta^{-1} \ln\beta & \text{if } x_b = 3/2\\
\beta^{-1} & \text{if } x_b > 3/2 \, ,
\end{array} \right.
\end{equation}
to $\mathcal{O}(\beta^{-1})$ in $\beta$ and $\mathcal{O}(e^{-2 \pi r / \beta})$ in $r / \beta$. 

The results in Eq. (\ref{simpbeta}) bear a close resemblance to the well-known expressions for the impurity specific heat $C_{imp}$ at criticality.\cite{AL2,FJ} In fact, as $C_{imp}$ is related to the thermodynamic impurity entropy $S_{imp}^{Th}$ via the relation $C_{imp} = -\beta \,\partial S_{imp}^{Th} / \partial \beta$, they describe the same power law, and one has the leading behavior
\begin{equation} \label{sth}
S_{imp}^{Th} = \ln g + \left\{
\begin{array}{ll}
\lambda^2 A_1 \beta^{2-2x_b} & \text{if } 1< x_b < 3/2\\
\lambda^2 A_2 \beta^{-1} \ln\beta & \text{if } x_b = 3/2\\
\lambda^2  A_3 \beta^{-1} & \text{if } x_b > 3/2 \, ,
\end{array} \right.
\end{equation}
as $\beta \to \infty$, where $A_1$, $A_2$ and $A_3$ are constants. Thus the von Neumann and thermodynamic impurity entropies have the same form on their leading scaling corrections in the limit $r/\beta \to \infty$ at low temperature. However, the amplitudes of these scaling corrections are different, as the von Neumann entropy acquires an additional amplitude factor from the mapping from the Riemann surface, not present in the thermodynamic entropy.

{\em Impurity entanglement entropy in Kondo systems.}
At zero temperature the von Neumann entropy measures the entanglement between the two parts of the bipartite system, and the impurity part $S_{imp}(r)$ in Eq.~(\ref{S_imp}) is then referred to as the impurity entanglement entropy.~\cite{SCLA}
We can therefore use the zero-temperature results in Eq.~(\ref{simp}) to predict the impurity entanglement entropy in various Kondo impurity models. They all share the common feature that the only significant zero-temperature length scale is the Kondo length $\xi_K\sim v_F / T_K$, the characteristic length scale at which screening is supposed to occur (here $v_F$ is the Fermi velocity and $T_K$ the Kondo temperature). In particular, the distance dependence of $S_{imp}$ can only come through $r/\xi_K$. Note that when the models describe two- (three-) dimensional systems the size $r$ of the block at the boundary will correspond to the radius of a disc (sphere) centered at the impurity (or the midpoint between the impurities when they are two). Thus the impurity entanglement entropy measures the impurity-generated entanglement between the part of the system within radius $r$ (including the impurity) and the rest of the system. It appears as a natural measure of the shape of this screening cloud since it captures the spatial distribution of the entanglement from the impurity. Compared to other ways of probing Kondo screening with entanglement,~\cite{BSB} the impurity entanglement entropy has the advantage of allowing analytical results based on BCFT. 

The BCFT approach,~\cite{Affleck} where the model is reduced to one spatial dimension with the impurity as a special boundary condition, is only valid in the limit $r \gg \xi_K$. Hence we expect the BCFT prediction for $S_{imp}$ at zero temperature to describe the large-distance decay of the Kondo screening cloud. By our exact analysis above, the leading $r$ dependence of $S_{imp}$ in Eq.~(\ref{simp}) has the same form as the leading $\beta$ dependence of $C_{imp}$, so the large-distance zero-temperature profile of the screening cloud is encoded by the impurity specific heat. The relation to the impurity entanglement entropy is less direct for the impurity susceptibility, which is governed by the same irrelevant boundary operator but through coupling to the bulk spin operator.~\cite{AL2} We now illustrate our result for a large class of Kondo impurity systems.

{\em The Kondo model.}
The original Kondo model describes a band of conduction electrons interacting with a single magnetic $s=1/2$ impurity at the origin:
\begin{equation} \label{kondo}
H_{K} = \hspace{-.1cm}\displaystyle \sum_{\vec{k},\alpha=\uparrow,\downarrow}  \hspace{-.2cm}\epsilon(\vec{k})  \psi^{\dagger \alpha}_{\vec{k}} \psi_{\vec{k} \alpha} + J \vec{S} \cdot \hspace{-.25cm}\sum_{\alpha, \beta = \uparrow,\downarrow} \hspace{-.2cm} \psi^{\dagger \alpha}(\vec{0}) \frac{\vec{\sigma}^{\beta}_{\alpha}}{2}\psi_{\beta}(\vec{0})\, . \ 
\end{equation}
At the low-temperature Kondo screening fixed point, $\ln g= 0$, and the leading irrelevant boundary operator is the stress-energy tensor $T$. As found in Ref. \onlinecite{SCLA}, adding the boundary perturbation $\delta H = - (\xi_K / 2) \,T(0)$ to the fixed point Hamiltonian $H_{ {\text BCFT}}$ gives $S_{imp}(r) = \pi \xi_K / (12 r)$ for $r \gg \xi_K$, in agreement with DMRG results and consistent with a valence-bond picture.

Breaking particle-hole symmetry introduces the exactly marginal charge current operator as a perturbation at the fixed point, only giving a constant shift of $S_{imp}(r)$.

{\em The two-impurity Kondo model.}
Adding a second spin-1/2 impurity to the Kondo model gives the two-impurity Kondo model (TIKM)
\begin{eqnarray} \label{HTIKM}
H_{{\text TIKM}} &=& \displaystyle \sum_{\vec{k},\alpha=\uparrow,\downarrow}  \epsilon(\vec{k})  \psi^{\dagger \alpha}_{\vec{k}} \psi_{\vec{k} \alpha}  + J \,[\  \vec{s}_c(\vec{r}_1) \cdot \vec{S}_1 
\nonumber \\
&& \quad +\, \vec{s}_c(\vec{r}_2) \cdot \vec{S}_2 \ ] + K \, \vec{S}_1 \cdot \vec{S}_2 \, ,
\end{eqnarray}
where $\vec{s}_c(\vec{r})= (1/2)\sum_{\alpha, \beta}\psi^{\alpha \dagger}(\vec{r})\vec{\sigma}^{\beta}_{\alpha} \psi_{\beta}(\vec{r})$. The BCFT solution of the model was found in Ref. \onlinecite{ALJ}. In short, the model features an unstable fixed point at $K = K_c \sim T_K$, where the system undergoes a quantum phase transition. At this fixed point $\ln g = \ln \sqrt{2} $, and the leading irrelevant boundary operator allowed by symmetry to appear as a perturbation is $L_{-1} \epsilon$, the Virasoro first descendant of the $\epsilon$ field, with scaling dimension $x_b=3/2$. However, being a Virasoro first descendant, $L_{-1} \epsilon$  will not give any contribution neither to the entanglement entropy nor any finite-temperature properties \textit{to any order} in perturbation theory. This follows from the evaluation of integrals of the type $\int_{-\infty}^{\infty} \, \textrm{d}\tau \langle L_{-1} \epsilon \rangle_{\mathcal{R}_n} = \int_{-\infty}^{\infty} \, \textrm{d}\tau \partial_{\tau} \langle \epsilon \rangle_{\mathcal{R}_n} $ which vanish due to the periodicity of the boundary of the  $n$-sheeted Riemann surface $\mathcal{R}_n$. The leading correction to the impurity entanglement entropy therefore comes from the stress-energy tensor $T$, precisely as in the single-impurity case. Thus, adding $T$ as a boundary perturbation to the fixed-point Hamiltonian $ H_{{\text BCFT}}$,
\begin{equation} \label{H_FP_TIKM}
H =  H_{{\text BCFT}} + a \sqrt{\xi_K} L_{-1} \epsilon (0) - b\, \xi_K T(0) \, ,
\end{equation}
with $a$ and $b$ dimensionless constants, gives the impurity entanglement entropy in the limit $r \gg \xi_K$ as
\begin{equation} \label{S_TIKM}
S_{imp}(r) = \ln \sqrt{2} + \pi b \, \xi_K / (6 r)\, .
\end{equation}

Breaking the SU(2) spin-rotational symmetry or the parity symmetry introduces a boundary operator with scaling dimension~\cite{ALJ} $x_b=3/2$. Particle-hole symmetry breaking can be either relevant or exactly marginal. Simultaneously breaking the particle-hole and parity symmetries adds another $x_b = 3/2$ boundary operator. \cite{MJ}

{\em The multichannel Kondo model.} The Hamiltonian is obtained by adding a channel index $i=1,2,...,k$ to the electrons in Eq.~(\ref{kondo}), so that the spin-$s$ impurity interacts with $k$ degenerate bands of conduction electrons. When $k > 2s$, the system is governed by an overscreened non-Fermi-liquid fixed point, corresponding to a boundary entropy \cite{AD} $\ln g = \ln \left[\sin(\pi(2s+1)/(2+k)) / \sin(\pi/(2+k))\right]$. The leading irrelevant boundary operator has scaling dimension \cite{AL2} $x_b=1+2/(2+k)$, in agreement with the original Bethe Ansatz solution. \cite{AD} Thus, when $k=2$ the distance-dependent part of $ S_{imp}(r)$ falls off as $\sim (\xi_K /  r) \ln (r / \xi_K)$. For arbitrary $k>2$, where $1< x_b < 3/2$, one gets 
\begin{equation} \label{S_MCKM}
 S_{imp}(r) = \ln g +  A \, (r / \xi_K )^{-4/(2+k)}, 
\end{equation}
for $r \gg \xi_K$, with $A$ a constant. We see that the more channels that are added, the more long-range entanglement appears to be generated from the screening of the impurity by the composite soliton-like objects formed in the non-Fermi liquid.~\cite{AD} The screening cloud falls off with an anomalous power law, a result rather similar to the one obtained by Barzykin and Affleck~\cite{BA} when defining it as the form of the equal-time spin-spin correlator.

Channel asymmetry is a relevant perturbation, \cite{ALPC} whereas particle-hole symmetry breaking is exactly marginal. \cite{AL2} The effect of spin-rotational symmetry breaking depends on $s$ and $k$; \cite{ALPC} however for the special cases where the perturbation is irrelevant the only difference is a change in the constant $A$ in Eq.~(\ref{S_MCKM})

Generalizing the multichannel Kondo model by extending the spin symmetry group from $SU(2)$ to $SU(N)$ gives the \textit{multichannel SU(N) Kondo model}. \cite{PGKS-ZJA} The leading irrelevant operator at the overscreened non-Fermi-liquid fixed point  now has scaling dimension $x_b = 1+N/(N+k)$. Hence
$ S_{imp}(r) = \ln g +  A \, (r / \xi_K )^{-2N/(N+k)}$ for $r \gg \xi_K$, with $A$ a constant and $\ln g$ depending on the particular representation of $SU(N)$.

{\em The two-impurity, two-channel Kondo model} features a continuous family of non-Fermi-liquid fixed points, \cite{GS} allowing for scaling-correction terms in $S_{imp}(r)$ from boundary operators with scaling dimensions ranging all the way from $x_b=1$ to infinity. 

The final case we consider is a \textit{Kondo impurity in a Luttinger liquid}. At zero temperature $\ln g = 0$, \cite{FN} but the leading perturbation typically has\cite{FJ} $1< x_b < 3/2$ and Eq.~(\ref{simp}) then predicts that $S_{imp}(r) \sim r^{2-2x_b}$.

{\em Discussion.}
We have provided a unified picture for the impurity entanglement entropy $S_{imp}$ for the large class of Kondo models in terms of the scaling corrections arising from the boundary operators of the corresponding BCFT, valid for $r \gg \xi_K$. When $r/\beta \to \infty$ the scaling part of the von Neumann entropy approaches the thermodynamic entropy,~\cite{CC1} and it has been argued that this is expected on general grounds.~\cite{SCLA} Our analysis shows that this reasoning can also be applied to the scaling corrections, but only in the following precise sense: The impurity von Neumann entropy and thermodynamic entropy have the same leading power-law dependence of $\beta$ at low temperatures, however with different amplitudes. This connects the exponents for the temperature scaling of the impurity specific heat with those for the large-distance scaling of the zero-temperature impurity entanglement entropy, through the von Neumann entropy. 

The result suggests that the large-distance profile of a zero-temperature Kondo screening cloud can be read off from the impurity specific heat at low temperature, an experimentally accessible observable. For the original single-impurity Kondo model, $S_{imp}(r)$ was indeed found~\cite{SCLA} to follow the same power law as the impurity specific heat $C_{imp}(\beta)$, showing that this also holds for the first-order correction from the stress-energy tensor. This is also what we see when considering the BCFT predictions for a number of more complex Kondo impurity models, thereby demonstrating the close connection between Kondo screening, entanglement and thermodynamics in quantum impurity systems.

{\em Acknowledgments $-$} We wish to thank Natan Andrei and Fabien Alet for helpful communications. This research was supported by the Swedish Research Council under Grant No. VR-2008-4358.

\end{document}